\documentclass[3p,times,twocolumn]{elsarticle}

\usepackage{ecrc}

\volume{00}

\firstpage{1}

\journalname{Nuclear Physics B Proceedings Supplement}

\runauth{}

\jid{nppp}

\jnltitlelogo{Nuclear Physics B Proceedings Supplement}

\usepackage{amssymb}
\usepackage{amsmath}

\usepackage[figuresright]{rotating}

\usepackage{color}

\begin{document}

\begin{frontmatter}



\dochead{}

\title{Boson-Jet Correlations in a Hybrid Strong/Weak Coupling Model for Jet Quenching in Heavy Ion Collisions}

\author[a]{Jorge Casalderrey-Solana}
\author[b]{Doga Can Gulhan}
\author[c,d]{Jos\'e Guilherme Milhano}
\author[a]{Daniel Pablos}
\author[b]{Krishna Rajagopal}

\address[a]{
Departament d'Estructura i Constituents
de la Mat\`eria and Institut de Ci\`encies del Cosmos (ICCUB),
Universitat de Barcelona, Mart\'\i \ i Franqu\`es 1, 08028 Barcelona, Spain
}

\address[b]{Laboratory for Nuclear Science and Department of Physics, Massachusetts Institute of Technology, Cambridge, MA 02139, USA}

\address[c]{CENTRA, Instituto Superior T\'ecnico, Universidade de Lisboa, Av. Rovisco Pais, P-1049-001 Lisboa, Portugal}
\address[d]{Physics Department, Theory Unit, CERN, CH-1211 Gen\`eve 23, Switzerland}
\begin{abstract}
We confront a hybrid strong/weak coupling model for jet quenching to data from LHC heavy ion collisions. The model combines the perturbative QCD physics at high momentum transfer and the strongly coupled dynamics of non-abelian gauge theories plasmas in a phenomenological way. By performing a full Monte Carlo simulation, and after fitting one single parameter, we successfully describe several jet observables at the LHC, including dijet and photon jet measurements. Within current theoretical and experimental uncertainties, we find that such observables show little sensitivity to the specifics of the microscopic energy loss mechanism. We also present a new observable, 
the ratio of the fragmentation function of inclusive jets to that of the associated jets in dijet pairs, which can discriminate among different medium models. Finally, we discuss the importance of plasma response to jet passage in jet shapes.
\end{abstract}


\end{frontmatter}


\section{Introduction}
\label{intro}
Jet quenching phenomena in heavy ion collisions involve processes occurring at a wide range of scales. The production and hard evolution of jets at high scales are well described by perturbative QCD. However, the interaction of the jet with QGP at a temperature not far above the transition temperature $T_c \sim \Lambda_{QCD}$ poses a serious challenge for current perturbative techniques. In fact, the collective properties of the plasma at this range of temperature are better described as a strongly coupled system than as a weakly coupled gas of quarks and gluons. It is then natural to expect non-perturbative physics to play an important role in the jet/plasma dialogue. The hybrid strong/weak coupling model developed in \cite{Casalderrey-Solana:2014bpa} takes into account these two facts by relying on holographic calculations for the description of this interplay, while treating the highly energetic parton branching perturbatively via DGLAP equations.

Currently there is no theoretical framework which can describe strong and weak coupling processes at different scales in a consistent manner. For this reason our model should be regarded as a phenomenological approach which exploits the big separation of scales from the virtuality to the temperature to combine the most relevant physical processes at each scale. 
Despite its simplicity, the model has proven to be a powerful tool in its confrontation with available measurements for various jet observables  \cite{Casalderrey-Solana:2014bpa,Casalderrey-Solana:2015vaa}, and in producing a broad range of definite predictions for LHC run II \cite{Casalderrey-Solana:2015vaa}.
In these proceedings we extend the comparison carried out in \cite{Casalderrey-Solana:2014bpa,Casalderrey-Solana:2015vaa} by both confronting the model with ATLAS jet data and exploring new sets of observables.

\section{A Hybrid Model}
Two key ingredients underlie the construction of our hybrid model. First, soft medium exchanges are assumed not to change the virtuality of the partons significantly and, consequently, jet evolution proceeds as in vacuum. We assign a lifetime to each of the constituents of the shower by a formation time argument. Second, we supplement the shower with an energy loss per unit path length analogous to that of a hard quark propagating in a strongly coupled plasma as described in \cite{Chesler:2014jva}. The rate of energy loss in this case is given by

\begin{equation}
\label{CR}
\frac{1}{E_{in}}\frac{dE}{dx}=-\frac{4x^2}{\pi x_{stop}^2 \sqrt{x_{stop}^2-x^2}} \,,
\end{equation}
 
 \begin{figure}[t]
\centering 
\includegraphics*[width=.5\textwidth]{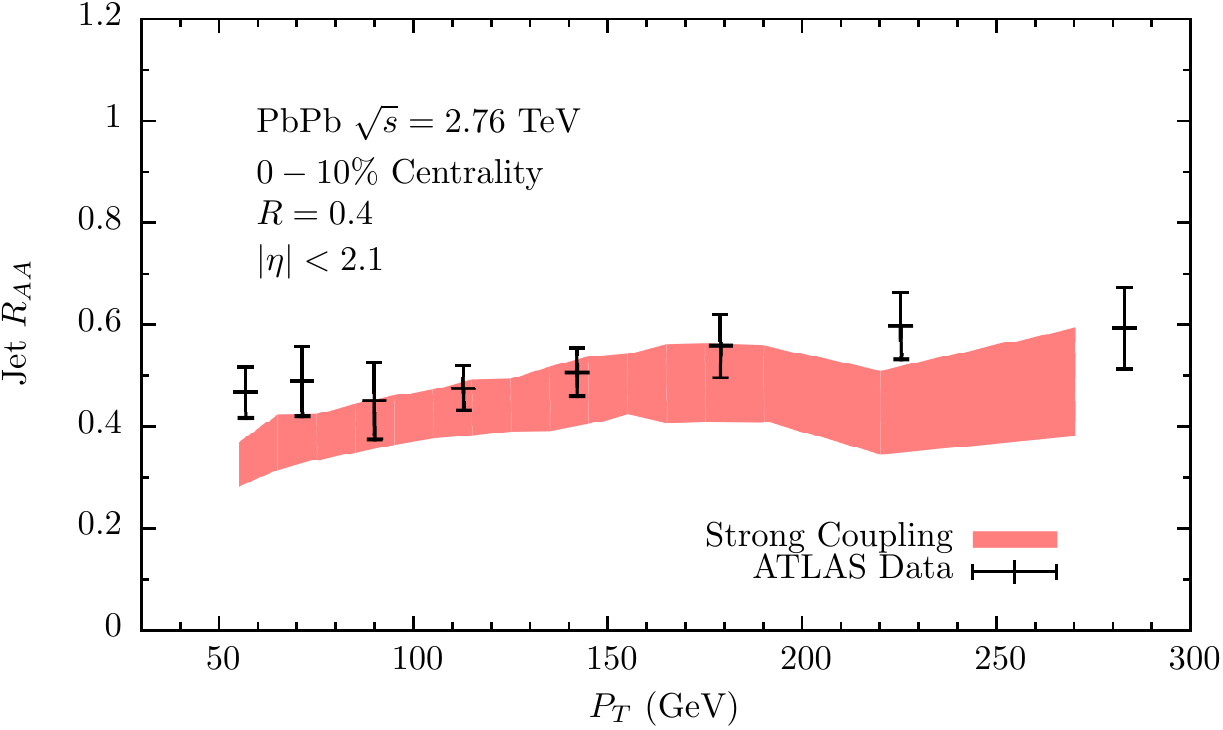}
\caption{Jet $R_{AA}$ for central events compared to ATLAS data \cite{Aad:2014bxa}. The single parameter in the model was fitted previously to CMS jet $R_{AA}$ data with different kinematical cuts~\cite{Casalderrey-Solana:2014bpa,Casalderrey-Solana:2015vaa}. 
The band represents a combination of theoretical and experimental uncertainty.}
\label{fig1} 
\end{figure}

\begin{equation}
x_{stop}=\frac{1}{2 \kappa_{SC}}\frac{E_{in}^{1/3}}{T^{4/3}} \, ,
\end{equation}
\noindent
with $x_{stop}$, the stopping distance of the parton, being the smallest matter thickness that results in a total parton energy loss.
Within the gauge/gravity duality context, the $E$ and $T$ dependence of this quantity has been shown to be robust. However, $\kappa_{SC}$ exhibits different parametric dependence for string based \cite{Chesler:2014jva} and wave-packet based \cite{Arnold:2010ir} computations. Nevertheless, for realistic values of the coupling, both these estimates for $\kappa_{SC}$ yield values of $\mathcal{O}(1)$. The model assumes that all the differences between the theories described by the duality and QCD can be absorbed in this quantity, which we consider to be a fitting parameter. Since the number of degrees of freedom of QCD is smaller than that of $\mathcal{N}=4$ SYM, we expect $\kappa_{SC}$ to be smaller than (but of order) $1$.

To check the sensitivity of the observables tested to the precise energy loss mechanism, whose parametric dependence can be encapsulated using an energy loss rate, we consider the control models
\begin{equation}
\bigg( \frac{dE}{dx} \bigg)_{rad}=-\kappa_R \frac{C_R}{C_F} T^3 x  \qquad \bigg( \frac{dE}{dx} \bigg)_{coll}=-\kappa_C \frac{C_R}{C_F} T^2 \, ,
\end{equation}

 \begin{figure}[t]
\centering 
\includegraphics*[width=.35\textwidth]{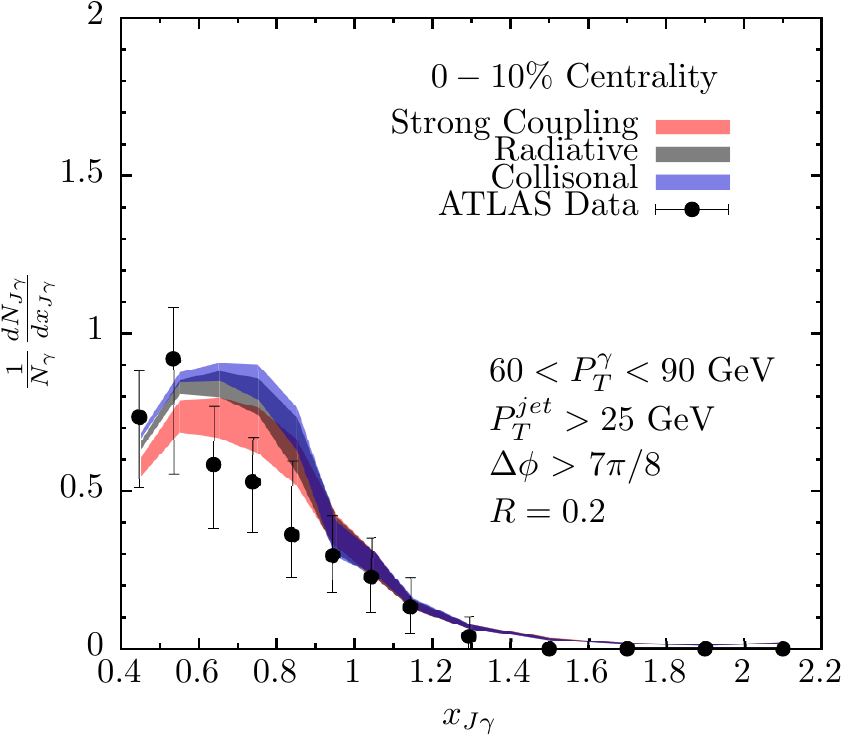}
\caption{Photon-jet imbalance for central events for our hybrid strong/weak coupling model as well as the two control models to ATLAS data \cite{ATLASphoton}.}
\label{fig2} 
\end{figure}
\noindent
where $\kappa_R$ and $\kappa_C$ are also fitting parameters. These models are inspired by radiative and collisional energy loss rates, but are not aimed at superseding more sophisticated implementations of these mechanisms. We use them only as benchmarks.

To compare to experiments we have implemented this model into a Monte Carlo. We simulate hard jet production and subsequent parton evolution through PYTHIA \cite{Sjostrand:2006za}. Hard jet events are embedded into a heavy ion environment by distributing the production point across the transverse plane according to an optical Glauber Monte Carlo, and using a realistic hydrodynamic profile \cite{Shen:2014vra} to simulate the evolution of the QGP that will be explored by the partons of the shower.

\section{Comparison with Data}
In previous publications~\cite{Casalderrey-Solana:2014bpa,Casalderrey-Solana:2015vaa} we fixed the fitting parameter $\kappa_{SC}$ by comparing our Monte Carlo simulations to CMS jet $R_{AA}$  data, for jets with $R=0.3$ lying within $100<p_T<110$ GeV for 0-10\% centrality class, obtaining 
the range $0.317 <  \kappa_{SC} < 0.452$. With the value of its single parameter fixed, the model is able to reproduce the energy and centrality dependence of CMS jet measurements of the nuclear modification factor, dijet imbalance, fragmentation functions, photon jet asymmetry and the suppression of jets produced in association with a photon~\cite{Casalderrey-Solana:2014bpa,Casalderrey-Solana:2015vaa}.

Here, we present the comparison of our model to ATLAS jet data on inclusive jet suppression and photon jet imbalance. In Fig.~\ref{fig1} we show our model computations for $R_{AA}$ for central events of jets with $R=0.4$ within $|\eta|<2.1$ as compared to ATLAS data~\cite{Aad:2014bxa}. Note that in the data points in Fig.~\ref{fig1} we add the systematic and statistical errors in quadrature. We have used the same value of the single parameter in the model that we obtained previously~\cite{Casalderrey-Solana:2015vaa} by fitting to CMS data even though here the jets are reconstructed with $R=0.4$ while in Ref.~\cite{Casalderrey-Solana:2015vaa} and
the CMS data $R=0.3$ is used. We see in Fig.~\ref{fig1} that the model compares well to data over a wide range of transverse momentum. In our current model implementation we have not included broadening effects, which may affect the comparison of jet suppression patterns at different values of $R$. We plan to address such effects in future work.

 \begin{figure}[t]
\centering 
\includegraphics[width=.5\textwidth]{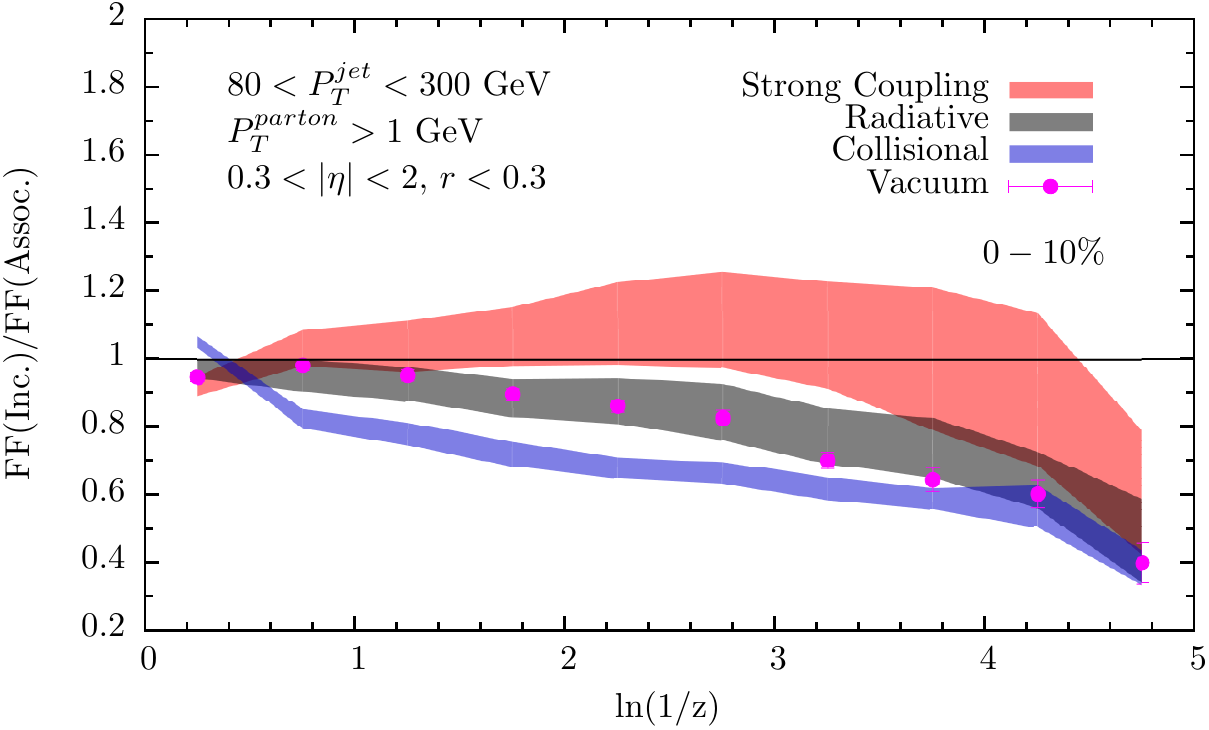}
\caption{Ratio of inclusive over associated jet fragmentation functions for PbPb at $\sqrt{s}=5.02$ ATeV. This observable is constructed with PbPb data only. The three bands represent the predictions obtained from the different models of the jet medium interaction.}
\label{fig3}
\end{figure}

In Fig.~\ref{fig2} we show the photon-jet imbalance distribution, namely the distribution of the variable $x_{J\gamma}\equiv \frac{p_T^{jet}}{p_T^{\gamma}}$ for jets with $R=0.2$ and in the pseudorapidity range $|\eta|<2.1$. Due to the important photon background in a heavy ion collision, one has to perform an isolation procedure to maximize the probability that the analyzed photon is actually prompt. The three bands represent the strong coupling energy loss model and the two control models. Even though there is some interesting separation among different models, within current uncertainties they are compatible with ATLAS data. The considerable increase in photon jet statistics expected for the upcoming LHC heavy ion run II may allow us to use this measurement to discriminate between different microscopic dynamics of the interaction
between the jet and the medium. (Detailed predictions for this observable at $\sqrt{s}=5.02$ ATeV can be found in \cite{Casalderrey-Solana:2015vaa}).

The small discriminating power of the inclusive results of \cite{Casalderrey-Solana:2014bpa} motivate us to explore intra-jet observables. Following 
Ref.~\cite{Casalderrey-Solana:2015vaa}, in Fig.~\ref{fig3} we propose a new observable. This is the ratio of the fragmentation functions of inclusive jets in PbPb events to those of  
associated jets (lower energy jets in a dijet pair) in PbPb collisions with the same kinematical cuts. The three bands correspond to the hybrid model and the two control models. For guidance, we have included the corresponding ratio for pp events. Unlike more inclusive observables, this ratio shows a clear separation among models. This separation is also clear in the intermediate $z$ region, which is less sensitive to effects affecting soft particles such as medium response. Note that this ratio confronts sets of more quenched and less quenched jets, and involves PbPb data only.

\begin{figure}[t]
\centering 
\includegraphics[width=.5\textwidth]{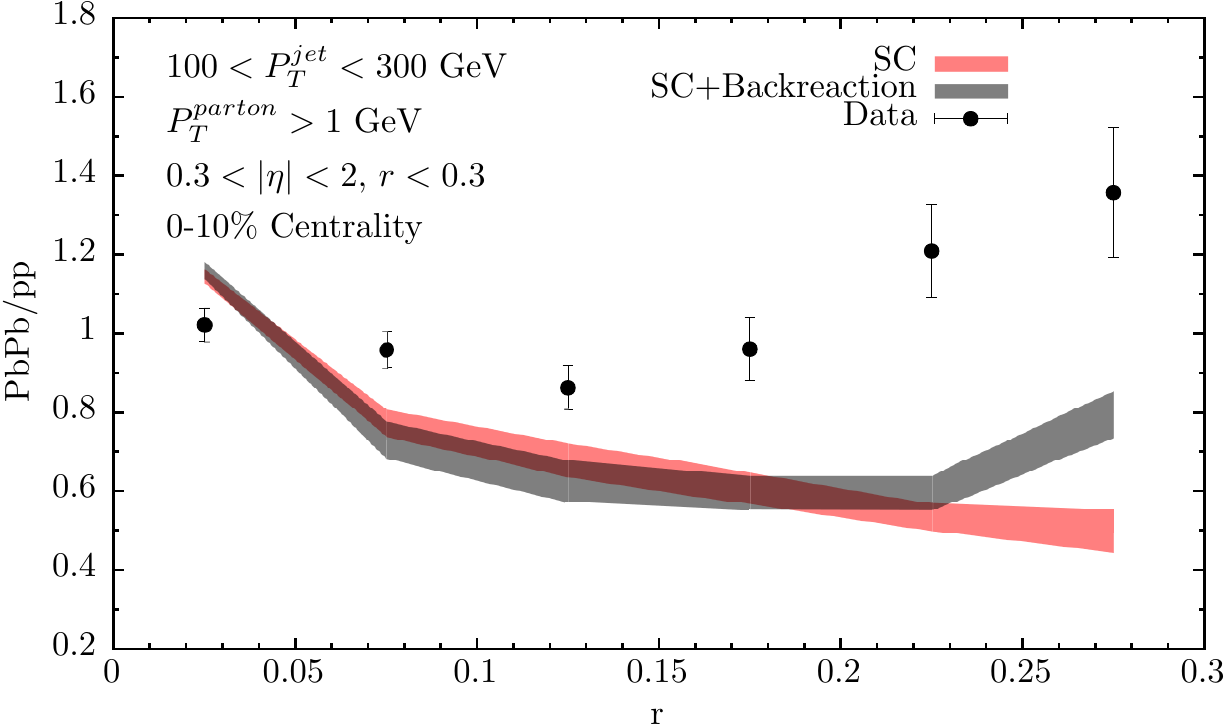}
\caption{Jet shapes ratio of strong coupling model in PbPb over pp compared to CMS data \cite{Chatrchyan:2013kwa}. The gray band represents a simple estimate of the effect of medium response to jet passage for this observable.}
\label{fig4}
\end{figure}

In Fig.~\ref{fig4} we compute another kind of intra-jet observable, the so-called jet shapes. The red band shows the results of our model calculations implemented as previously described. This implementation fails to reproduce the enhancement of in-medium jet shapes at large $r$ observed in CMS data \cite{Chatrchyan:2013kwa}. However, since highly energetic vacuum jets are known to have only around the 2\% of their energy at the periphery, it is expected that small amounts of extra energy deposited in such bins in PbPb events could have important effects on the ratio. One natural process able to produce this effect is the backreaction of the medium to the energy deposited by the jet, which makes the QGP fluid locally a little hotter and faster. We estimate such a medium response effect on the QGP by assuming small perturbations on top of a boost invariant fluid. We also consider that the fluctuations are narrow in rapidity around the jet direction. By making use of the Cooper-Frye formula, we derive the one body distribution for the extra emission of thermal particles off a boosted, heated up fluid cell:
\begin{equation}
\begin{split}
& E \frac{dN}{d^3p}=\frac{1}{32 \pi}\frac{m_T}{T^5}\textrm{cosh}(y-y_j)e^{-\frac{m_T}{T}\textrm{cosh}(y-y_j)}\\
&\big[ p_T\Delta P_T \textrm{cos}(\phi-\phi_j)+\frac{1}{3}m_T\Delta M_T\textrm{cosh}(y-y_j) \big],
\end{split}
\end{equation}
\noindent
where $\Delta P_T$ and $\Delta M_T$ are the lost transverse momentum and mass by the jet respectively, with $y_j$ its rapidity and $\phi_j$ its azimuthal angle. Note that negative contributions are present at  large azimuthal separation; these diminish the background contribution for that region in space. Event by event, we fully specify the distribution of extra particles via four momentum conservation using a Metropolis algorithm. The implications of this estimation can be seen in Fig.~\ref{fig4}. We observe an important enhancement on the tail of the jet shapes ratio. Although this simple estimate cannot reproduce the measured enhancement, this observation shows that the description of medium response is important to understand this data. This effect has also been studied in \cite{He:2015pra}, obtaining similar conclusions.

\section{Conclusions}
The combination of these proceedings and  Refs.~\cite{Casalderrey-Solana:2014bpa,Casalderrey-Solana:2015vaa} shows that our phenomenological hybrid model is capable of describing a wide set of jet observables, all after fitting a single parameter. The extracted value implies that the stopping distance of a QCD plasma is 3$\sim$4 times larger than that of a $\mathcal{N}=4$ plasma at infinitely strong coupling with the same temperature,  consistent with the smaller number of degrees of freedom.

 Inclusive jet observables do not show strong discriminating power on the underlying dynamics. As an example, the difference in photon jet imbalance results shown in these proceedings are smaller than current data uncertainties, thus preventing us from extracting a stark conclusion about the nature of the degrees of freedom of the QGP plasma. Future LHC run II data may provide sufficient discriminative power. More differential observables also contribute to discrimination. In particular, the ratio of inclusive over associated fragmentation functions amplifies the difference in path length and energy dependence of the different models explored. It would be interesting to measure this observable in the upcoming LHC run.
 
 We have also shown how the thermalized energy extracted from the high energy jet can translate into an enhancement of soft tracks with respect to the unperturbed hydrodynamical background, leading to an important modification of the tail of the jet shapes distribution. It seems unavoidable that such effects are essential in order to accurately reproduce jet quenching phenomena.

\vspace{0.4cm}
\noindent{\bf Acknowledgements}
We thank A. Angerami for useful discussions and correspondence regarding the comparison of our model to ATLAS data.
The work of JCS was  supported by a Ram\'on~y~Cajal fellowship.  The work of JCS and DP was 
supported by  the  Marie Curie Career Integration Grant FP7-PEOPLE-2012-GIG-333786, by grants FPA2013-46570 and  FPA2013-40360-ERC
and MDM-2014-0369 of ICCUB (Unidad de Excelencia `Mar\'ia de Maeztu') 
 from the Spanish MINECO,  by grant 2014-SGR-104 from the 
Generalitat de Catalunya 
and by the Consolider CPAN project. 
The work of DCG and KR was supported by the U.S. Department of Energy
under Contract Numbers DE-SC0011088 and DE-SC0011090, respectively.
The work of JGM was supported by Funda\c{c}\~{a}o para a Ci\^{e}ncia e a Tecnologia (Portugal) under  project  CERN/FP/123596/2011 and contract `Investigador FCT -- Development Grant'.



\nocite{*}
\bibliographystyle{elsarticle-num}
\bibliography{Pablos_D}







\end{document}